\documentclass[namedreferences]{SolarPhysics}
\usepackage[optionalrh]{spr-sola-addons}
\usepackage{graphicx}
\usepackage{color}                       
\usepackage{url}

\begin{document}    
\begin{article}
\begin{opening}
\title{Modelling of EIS Spectrum Drift from Instrumental Temperatures}
\author{S.~\surname{Kamio}$^{1}$\sep
        H.~\surname{Hara}$^{2}$\sep
        T.~\surname{Watanabe}$^{2}$\sep
        T.~\surname{Fredvik}$^{3}$\sep
        V. H.~\surname{Hansteen}$^{3}$
       }
%

%
\institute{$^{1}$ Max-Planck-Institut f\"ur Sonnensystemforschung (MPS), 37191 Katlenburg-Lindau, Germany
email: \url{skamio@spd.aas.org}\\
$^{2}$ National Astronomical Observatory of Japan, Mitaka, Tokyo 181-8588, Japan\\
$^{3}$ Institute of Theoretical Astrophysics, University of Oslo, P.O. Box Blindern, N0315 Oslo, Norway
           }

\begin{abstract}
An empirical model has been developed to reproduce the drift of the spectrum
recorded by the EIS on {\it Hinode} using instrumental temperatures and
relative motion of the spacecraft.
The EIS spectrum shows an artificial drift in wavelength dimension
in sync with the revolution of the spacecraft, which is
caused by temperature variations inside the spectrometer.
The drift amounts to $70$~km~s$^{-1}$ in Doppler velocity
and introduces difficulties in velocity measurements.
An artificial neural network is incorporated to establish
a relationship between the instrumental temperatures and the spectral
drift. This empirical model reproduces observed spectrum shift
with an rms error of 4.4~km~s$^{-1}$.
This procedure is robust and applicable to any spectrum
obtained with EIS, regardless of of the observing field.
In addition, spectral curvatures and spatial offset
in the north -- south direction are determined to compensate for
instrumental effects.
\end{abstract}
\end{opening}

\section{Introduction}
The EUV Imaging Spectrometer (EIS; \opencite{culhane2007}, \opencite{korendyke2006}, \opencite{lang2006}) on {\it Hinode} \cite{kosugi2007}
is a powerful instrument for plasma diagnostics in the solar corona.
The spectrometer can obtain a lot of EUV spectral lines emitted from
a wide temperature range of plasma and allows us to study
the dynamics of the solar corona.
The spectrometer is capable of measuring spectrum shift
in prominent emission lines with an accuracy better than 3~km~s$^{-1}$.
However, one of the challenges in EIS data analysis is
that the spectrum shows a quasi-periodic shift in the wavelength
dimension.
Since the shift is synchronized with the spacecraft
revolution around the Earth, it must be
an instrumental effect.

The artificial spectral drift amounts to $70$~km~s$^{-1}$ in
Doppler velocity scale and introduces a significant effect on
 velocity measurements.
A commonly used method so far is to assume
that a net Doppler shift vanishes in a quiet region.
As a quiet region is not always included in the EIS field of view,
the assumption of zero net velocity is not applicable in some cases.
It has been reported that the spectrum drift is connected with
temperature variations in the spectrometer.
\inlinecite{brown2007} showed that the spectral drift is
correlated with grating temperature variation.
But their relationship is not simple because
a slight phase difference is found between
them.
A qualitative comparison between the spectrum drift
and instrumental temperatures shows that the
spectral drift reflects all aspects of
the temperature variations.
\inlinecite{rybak1999} demonstrated that 
periodic spectral shifts of the SUMER spectrometer can be
corrected by using instrumental temperatures.
However, a more complex method is needed to reproduce
the spectral drifts of the EIS.

Our goal is to develop an empirical model to reproduce
spectral drift from instrumental temperatures
so that the Doppler velocity can be determined from
any spectra obtained with EIS.
The organization of the paper is as follows:
The scheme of the model and definition of their input and output are
described in Section 2.
A comparison between the model and the measured spectral position is
presented in Section 3. 
In Section 4, the performance of the model and its
application to observations are discussed.
The curvatures and spatial offset
of the spectrum, which are essential for data analysis with EIS, 
are described in the Appendices.
\section{Method}

\begin{figure}
\centerline{\includegraphics[width=0.8\textwidth,clip=]{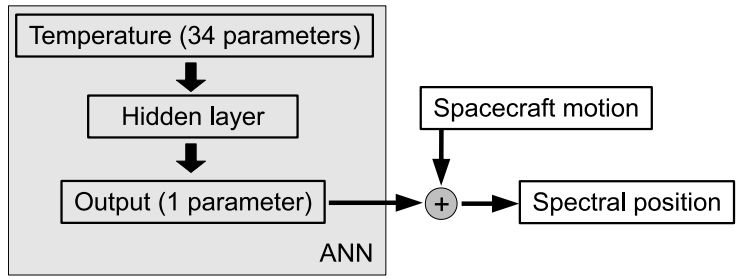}}
\caption{A schematic diagram of the model to estimate spectral position on the detector from instrumental temperatures and relative motion of the spacecraft.}
\label{fig:diagram}
\end{figure}

Figure \ref{fig:diagram} shows a schematic diagram of the
model to reproduce spectral position on the detector
from instrumental temperatures.
Its input consists of instrumental temperatures of the EIS.
The EIS employs 31 temperature sensors which allows
measurement of the temperature distribution of the optical bench.
They are monitored at 10 s intervals by the instrument.
These instrumental temperatures show seasonal variation
caused by the variation of the Sun -- Earth distance.
In addition, some temperatures exhibit short term variation
in sync with the orbital revolution of the spacecraft (98 min period),
while the rest of temperatures show only seasonal variations.
For the short varying group,
temperatures at 5 min earlier and later are also incorporated
into the model to account for their time derivatives.
Temperature sensors placed close to each other are redundant,
and are reduced to one temperature to minimize the number of inputs.
The total number of inputs for the model is 34.
 
An artificial neural network (ANN) is employed
to deduce spectral drift from the instrumental temperatures.
It has been demonstrated that the ANN is an efficient method
for solving a complex problem,
such as deriving atmospheric parameters from Stokes profiles
\cite{carroll2001,socasnavarro2003,borrero2010}.
The basics of the ANN are described in \inlinecite{bishop1995}.
The ANN is realized as layers of nodes, which is called
a feed forward network.
Two nodes in neighbouring layers are connected by a link
which has a weight and a bias to define a relationship
between the nodes.
The ANN consists of input layer, hidden layer,
and output layer.
In our model, the input layer is a set of instrumental
temperatures.
The hidden layer is a layer of nodes
connecting the input and the output.
A single hidden layer has been employed to keep
the model simple.
The output is a single value representing the
spectral position under given instrumental temperatures.
The ANN is trained to reproduce the relationship
between input and output by optimizing weights and biases
for the links.
The ANN allows us to deduce an empirical relationship
between instrumental temperatures and spectral drift
from the huge data set.

It is necessary to compensate for the Doppler shift caused by
the spacecraft motion relative to the observed object
on the Sun.
The motion of the spacecraft is calculated onboard to
tune the narrow band filter of the Solar Optical Telescope
\cite{tsuneta2008} on {\it Hinode}.
The relative velocity of the spacecraft consists of the spacecraft revolution,
the Sun -- Earth distance variation, and solar rotation of the observing region
which add up to $\pm6$~km~s$^{-1}$ at the largest.
The calculated velocity is retrieved from the spacecraft status data
and is added to the output of the ANN to estimate spectral position
on the detector
(Figure \ref{fig:diagram}).

\begin{figure}
\centerline{\includegraphics[width=0.8\textwidth,clip=]{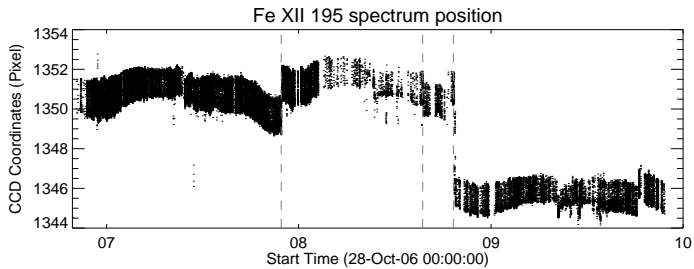}}
\caption{Temporal evolution of spectral position of Fe~{\sc xii}
$\lambda$195.12~\AA~ on the detector.
Vertical dashed lines indicates times of discontinuity in the spectral position.}
\label{fig:fullterm}
\end{figure}

The output of the model is a spectral center position
in Fe~{\sc xii} $\lambda$195.12~\AA .
This emission line is one of the EIS core lines,
hence is included in all observations.
As it is an intense emission line, the spectrum center can be accurately
determined in the quiet region as well as in the active region.
To minimize the effect of localized flow on the Sun,
data obtained with a slit length of 240 arcsec or greater
are selected for this analysis.
All EIS data meeting the criteria are retrieved from
{\it Hinode} archive in DARTS \cite{matsuzaki2007}.
A Gaussian fit is applied to the spectrum to determine the
center position at each slit height.
The instrumental spectral curvature is corrected by subtracting
it from the measured spectral center position
(see Appendix A for the spectral curvature of the EIS).
Then mean spectral position at each exposure
is derived from an averaged spectral center position
along the slit.

Figure \ref{fig:fullterm} presents a temporal variation of
the spectral position for three years since the {\it Hinode} launch.
A short term variation caused by the spacecraft revolution,
which is observed throughout the period,
amounts to two pixels in peak to peak amplitude.
A seasonal variation of the spectral position
is primarily due to the variation of Sun -- Earth distance
in one year cycle.
The eclipse season in May -- July and
heater configuration changes also affect the spectral position.
Two noticeable discontinuities in spectral position are
found at the time of slit focus adjustment on 24 August 2008
and grating focus adjustment on 21 October 2008,
which are marked by dashed lines in Figure \ref{fig:fullterm}.
A large discontinuity on 29 November 2007, which is indicated
by dashed line in Figure \ref{fig:fullterm}, is attributed to
a heater configuration change.

An ANN was built by using the multilayer perceptron software
package provided in the ROOT data analysis framework \cite{brun1997}
\footnote{See also \url{http://root.cern.ch/}.}.
The package is an open source software written in C++ language.
Because the ANN is an empirical method,
it is important to prepare a large number of data sets so
that they represent all possible conditions of the instrument.
To reproduce seasonal variation of the spectral position,
the ANN needs to be developed from data set 
at least for one year.

Series of data are divided at the times of
the grating and the slit movements,
which changed the optical alignment
of the spectrometer.
On top of these discontinuities,
data are also separated at the heater configuration change on
29 November 2007 because the discontinuity is too large
to be reproduced by a single model.
Three independent models A, B, and C are developed for
the split time series,
which are given in the second column of Table \ref{table:model}.
No independent model is built between 24 August 2008 and 20 October 2008
because of the small number of samples during that short interval.

The ANN is trained using the back propagation algorithm
provided with the multilayer perceptron package of the ROOT.
The performance of our model is measured by a residual
between measured spectral position and the model output.
The major causes of the residual are the error of the ANN
and flows on the Sun which are independent of
instrumental temperatures.
Since we employ a large number of data, velocities of the flows are
assumed to be random.
Thus, the residual is regarded as the upper limit of the ANN error.
In the following section, root mean square (rms) of the residual
is used as a measure of the ANN error.
The data set for each model is randomly divided into 80\%~of training
group and and 20\%~of test group.
In the course of the training, the ANN is trained with the training group,
while the performance of ANN is evaluated with the test group.
The training process is repeated until the performance reaches
a steady state.

A byproduct of our work is determination of the offset between
the $1''$ wide slit and $2''$ wide slit of EIS.
In principle, spectral drift caused by instrumental temperature variations
should work in the same way for both slits.
Two independent models are developed for $1''$ slit and $2''$
slit spectra in the period A. The offset between the $1''$ slit and the $2''$ slit
in the wavelength dimension has been derived from the difference
between two models, which is 8.20 in CCD pixels.
Spectra recorded with the $1''$ slit and the $2''$ slit of EIS are
merged after compensating for the offset between two slits.
To increase the number of data set, the merged data are used
for building three ANNs.
The composite numbers of $1''$ data and $2''$ data
are presented in the third column of Table \ref{table:model}.

Finally, the gap between the model B and C is filled by
extending the model C, assuming that the grating adjustment
on 21 October 2008 produced only a constant offset in spectral position.
The spectral positions between 24 August 2008 and 21 October 2008
are reproduced the best when the model C is shifted by 4.88 pixel.
For this period, a modified model C' is composed by adding a constant
offset of 4.88 pixel to model C.

\section{Results}
\subsection{Performance of the Model}
\begin{table}
\caption{Parameters for the models}
\label{table:model}
\begin{tabular}{ccccc} 
\hline                   
Model & Period & Samples & \multicolumn{2}{c}{rms of residual}\\
& & & (pixel) & (km~s$^{-1}$) \\
\hline
A & 03 Nov 2006 -- 28 Nov 2007 & $6.5\times10^5$ & 0.12 & 4.0 \\
B & 29 Nov 2007 -- 23 Aug 2008 & $9.5\times10^4$ & 0.16 & 5.4 \\
C' & 24 Aug 2008 -- 20 Oct 2008 & $1.1\times10^4$ & 0.14 & 4.7 \\
C & 21 Oct 2008 -- 28 Nov 2009 & $1.6\times10^5$ & 0.15 & 5.0 \\
\multicolumn{2}{c}{Total} & $9.2 \times10^5$ & 0.13 & 4.4\\
 \hline
\end{tabular}
\end{table}

Table \ref{table:model} presents the parameters for the developed models.
The last two columns are the rms of the residuals,
which are difference between the measured spectral position and
the model estimation.
The residual spectral shift is also converted to Doppler velocity at
Fe~{\sc xii} $\lambda$195.12~\AA .
Model A, which is created from the largest number of samples
gives the lowest residual.
The performance of model B is worse than the others,
which is probably due to a small number of samples
and its period falling short of one year.
Model C', which is built by shifting model C,
gives a similar residual to the other models.
The last row of Table \ref{table:model} presents the overall performance
of the composite model.
The composite model reproduces the spectral drift of the EIS with
a moderate accuracy of 4.4 km~s$^{-1}$ all through the period.

\begin{figure}
\centerline{\includegraphics[width=0.8\textwidth,clip=]{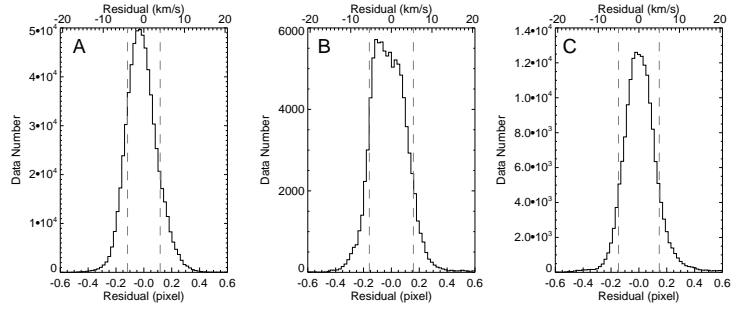}}
\caption{Distributions of the residuals for the three models. Dashed lines indicate the rms of the residuals.}
\label{fig:dist}
\end{figure}

Figure \ref{fig:dist} shows distributions of the residual for
the three models. Vertical dashed lines in each plot mark the
rms of the residual, which are also presented in
Table \ref{table:model}.
The distributions show that the majority of the residuals fall within
the rms ranges, but the residuals have greater values
in some cases.

\begin{figure}
\centerline{\includegraphics[width=0.8\textwidth,clip=]{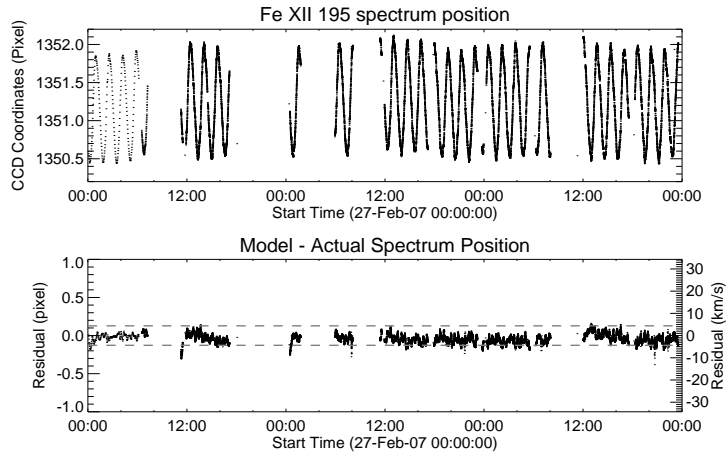}}
\caption{Top: Temporal evolution of spectral position Fe~{\sc xii} $\lambda$195.12~\AA~in CCD pixel coordinates. Bottom: difference between the model estimation and measured spectral positions in unit of pixel coordinate. Right vertical axis indicates
Doppler velocity scale. Dashed lines indicates $\pm 1\sigma$ levels of the residual, which is $\pm0.13$~pixel.}
\label{fig:day}
\end{figure}

Figure \ref{fig:day} presents a time series of measured spectral positions
and residuals in a day period, when the spacecraft
does not experience an eclipse.
The spectral position exhibits sinusoidal variation
with the orbital period of the spacecraft, which amounts to 1.5 pixels
in peak to peak amplitude.
In the bottom plot, the residuals are virtually suppressed below
$\pm0.13$~pixel range,
which means that the model reproduces the sinusoidal variation caused by
instrumental temperatures quite well.

\begin{figure}
\centerline{\includegraphics[width=0.8\textwidth,clip=]{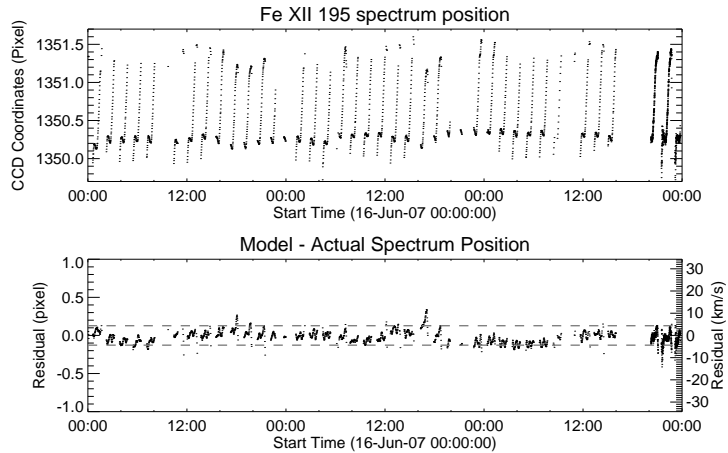}}
\caption{The same as Figure \ref{fig:day} but for an eclipse period.}
\label{fig:ecl}
\end{figure}

Figure \ref{fig:ecl} presents a time series in an eclipse period,
when the spacecraft experiences an eclipse in each orbital revolution.
The variation of spectral position is no longer sinusoidal but
a peculiar curve with a large amplitude of two pixels.
This is primarily because the EIS instrument experiences
large temperature variation in the day -- night cycle of the spacecraft.
But the residuals in the lower plot still remain at the same level as the day period,
which demonstrates that the model can also work in an eclipse period.

\subsection{Comparison with Other Correction Methods}

\begin{figure}
\centerline{\includegraphics[width=0.8\textwidth,clip=]{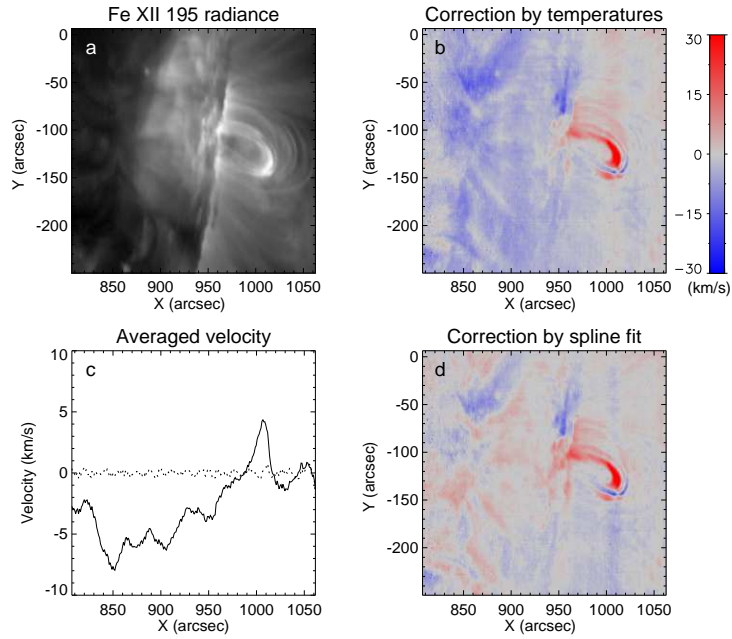}}
\caption{Coronal loops observed after a flare at the West limb on 17 Dec 2006.
a: The emission  in Fe~{\sc xii} $\lambda$195.12~\AA~ displayed in log scale.
b: Doppler velocity derived from Fe~{\sc xii} $\lambda$195.12~\AA~
after correction in the present paper.
c: Profiles of averaged velocity at each exposure.
Solid line and dotted line indicate resultant velocity after correction
by the method in the present paper and the spline fit.
d: Doppler velocity after subtracting spline fit to the observed spectral position.}
\label{fig:flare}
\end{figure}

Figure \ref{fig:flare} shows coronal loops observed after a flare on
17 December 2006 which was studied by \inlinecite{hara2008}.
Figure \ref{fig:flare}b shows the Doppler velocity after applying
the correction method developed in the present paper.
The velocity averaged at each exposure is plotted in Figure \ref{fig:flare}c
by solid line.
For comparison, velocity derived from
a cubic spline fit to the spectral drift, which is implemented in the
{\tt eis\_orbit\_spline} procedure in the EIS tree of Solar
Software\footnote{Solar Software (SSW) is available at \url{http://www.lmsal.com/solarsoft/}.},
is also presented in Figure \ref{fig:flare}d.

A major difference between the two methods is
that the spline fit method, which has been commonly used for
velocity measurement, assumes the mean velocity
in the observed field should be zero,
while our method does not assume that.
Our model estimates spectral drift only from instrumental
temperatures and relative motion of the spacecraft.
Averaged velocities plotted in Figure \ref{fig:flare}c characterize
two methods;
the velocity determined by spline fitting (dotted line) exhibits zero velocity
all through the raster scan, while the average velocity
processed by our method (solid line) deviates from zero.
However, the assumption of zero velocity might not be
valid in active regions, where localized high speed flows
are commonly detected.
One of major advantages of our method is that the resultant velocity is not
affected by an average velocity in the observing field.

\subsection{Extending the Method to Other Wavelengths}
\begin{figure}
\centerline{\includegraphics[width=0.8\textwidth,clip=]{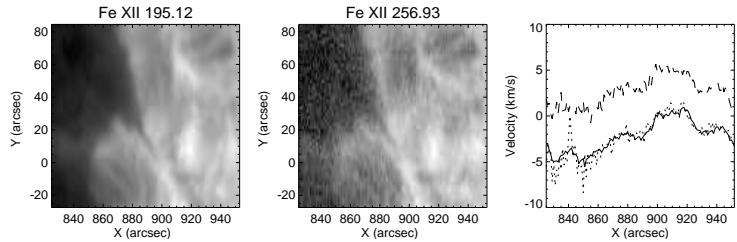}}
\caption{Left: A cut-out of an active region observed on 24 May 2007. The emission  in Fe~{\sc xii} $\lambda$195.12~\AA~ is displayed in log scale.
Middle: The emission in Fe~{\sc xii} $\lambda$256.93~\AA .
Right: Solid line and dotted line respectively show
averaged Doppler velocities at each exposure in Fe~{\sc xii} $\lambda$195.12~\AA~ and in Fe~{\sc xii} $\lambda$256.93~\AA.
Dashed line indicates the velocity for Fe~{\sc xi} $\lambda$180.40~\AA , which
is shifted by +5~km~s$^{-1}$ to avoid overlap.
}
\label{fig:ar}
\end{figure}
Although this correction method is developed for the Fe~{\sc xii}
$\lambda$195.12~\AA~ emission line, the instrumental drift of the spectrum should
also work in the same way for other wavelengths.
The validity of our correction method is examined by using 
Fe~{\sc xii} $\lambda$195.12~\AA~ in the short wavelength band and
Fe~{\sc xii} $\lambda$256.93~\AA~ in the long wavelength band.
As the two lines are emitted from the same type of ion,
they must indicate identical Doppler velocities.
These two lines allow us
to asses if the correction method derived from
the short wavelength band can be also applied to
the long wavelength band.

Figure \ref{fig:ar} shows a part of an active region observed in
Fe~{\sc xii} $\lambda$195.12~\AA~ and $\lambda$256.93~\AA~
on 24 May 2007.
Correction of the spectral drift is mandatory
because it took 60 min to complete the raster scan.
To compare the two emission lines, a spatial offset along the EIS slit
must be corrected, which is about 16 arcsec in the north -- south direction
(see Appendix B for the spatial offset of the spectrum).
Then the estimated spectral drift on the detector is subtracted
from both lines.
It must be noted that the instrumental spectral shifts are not in
the velocity scale
but in the detector pixel scale, because the velocity scale varies with
wavelength;
Doppler velocities corresponding to one pixel width of the detector are
34~km~s$^{-1}$ at $\lambda$195.12~\AA~
and 26~km~s$^{-1}$ at $\lambda$256.93~\AA .
Two panels in Figure \ref{fig:ar} present radiance maps after correcting
the spatial offset.
They show quite similar structure in the active region, 
although Fe~{\sc xii} $\lambda$256.93~\AA~ appears
noisier because of its low emission. 
Averaged Doppler velocities at each exposure is plotted in
the right panel of Figure \ref{fig:ar}.
The velocities inferred from two Fe~{\sc xii} lines are in good
agreement, which proves that our method of spectral
drift correction can be applied to the long wavelength band
as well as to the short wavelength band.
An increased deviation in the left part of the plot is
due to a large uncertainty caused by a low emission
in Fe~{\sc xii} $\lambda$256.93~\AA.
In addition, the velocity derived from Fe~{\sc xi} $\lambda$ 180.40~\AA~
is displayed in the right panel with +5~km~s$^{-1}$ offset.
Although it is emitted from a different ionization level,
the velocity profile is more or less the same as those of
Fe~{\sc xii} emission lines.

\section{Discussion}

We constructed an empirical method to compensate for
the spectral drift from instrumental temperature changes and the
motion of the spacecraft.
It consists of three different ANNs to accommodate
discontinuities in spectral positions.
The composite model reproduces observed spectral
drift with moderate accuracy; rms of the residual is
0.13~pixel or 4.4~km~s$^{-1}$ at Fe~{\sc xii} $\lambda$195.12~\AA .
The results demonstrate that  the spectral drift of
two-pixel amplitude is effectively suppressed by this method,
which proves that the main cause of spectral shift is
temperature variations in the spectrometer and
relative motion of the spacecraft.

Incompleteness of the model is considered as a cause of
the residuals between the actual spectral positions and the estimations.
Although the ANNs are built from 34 temperature inputs,
they may not be sufficient to represent
the temperature distribution of the spectrometer completely.
In addition, short exposure data are incorporated
for the sake of a better coverage of observing period,
which has a large uncertainty in spectral position
because of low counts.
Considering that the accuracy of a Gaussian fit
is roughly 0.1 pixel for a typical spectrum recorded by EIS,
the performance of 0.13 pixel uncertainty is reasonable.

Another possible cause of the error is due to flows in
the solar corona.
Even though the spectral positions are averaged
over 240 arcsec of the slit, high speed flows in
active regions could alter the spectral position.
However, the majority of spectra have been
obtained in quiet regions because of the
low solar activity in 2006 -- 2009.
Thus, active regions affect only a small
fraction of the data.

As the ANN is optimized to minimize the discrepancy
between the model estimate and measured spectral position,
it assumes that mean velocity of the entire data set
is zero. 
Since the large number of data sets used in
building the model virtually covers the entire Sun,
the velocity origin should be adjusted to the global average
of the Fe~{\sc xii} $\lambda$195.12~\AA~ emission line.
It has been reported in the literature that
the net velocity of coronal emission lines
deviates from zero \cite{peter1999,teriaca1999}.
\inlinecite{peter1999} determined a net blueshift of
$4.5\pm1.3$~km~s$^{-1}$
in Mg~{\sc x} formed at $10^6$~K at the disk center.
\inlinecite{teriaca1999} found a blueshift of
$9.8\pm1.6$~km~s$^{-1}$ in Fe~{\sc xii} $\lambda$1242.0~\AA~
in an active region.
If net upflow exists in the corona,
observed spectra should exhibit a center to limb
variation of the Doppler shift, which
could partly account for the uncertainty of our velocity correction
method.
However, a comparison between mean
velocities at the disk center and at the limb
in our data indicates no significant difference.
It is fair to claim the amount of center to limb
variation is within the uncertainty of the correction
method or 4.4 km~s$^{-1}$,
as this correction method may not be accurate
enough to measure a small velocity.

The velocity calibration method developed in the present paper
is suitable for active region studies, as it does not require
a velocity reference in the observing field.
A commonly used method so far is to assume that the average
velocity at each slit position is zero.
But the mean velocity in active regions may deviate from zero
since localized high speed flows are frequently observed.
For example, the coronal loops in Figure \ref{fig:flare}
exhibits a significant Doppler shift of about 40~km~s$^{-1}$,
which was interpreted as siphon flow along the loops \cite{hara2008}.
\inlinecite{harra2008} found a steady outflow up to 50~km~s$^{-1}$
in the periphery of an active region, where magnetic
fields extend to the outer corona.
A 15 km~s$^{-1}$ blueshift found at around $X = 850$ in Figure \ref{fig:flare}
could be interpreted as an outflow in the vicinity of the active region.
The correction by the zero velocity assumption
could be improved by carefully selecting a quiet region in the
observing field, but a quiet
region is not always available in active region observations.
The advantage of our method is that it does not
rely on the mean velocity in the observing field
and is applicable to any observations with the EIS.

\section{Conclusions}

An empirical method for velocity calibration is constructed using data from
three years of observations with the EIS.
It estimates instrumental spectral drift from temperatures inside
the spectrometer and relative motion of the spacecraft.
The model reproduces observed spectral position
on the detector with a moderate accuracy of 0.13 pixel
or 4.4~km~s$^{-1}$ at the Fe~{\sc xii} $\lambda$195.12~\AA~ emission line.
It proves that the spectral drift, which introduces difficulties in velocity
measurement with EIS, is primarily caused
by temperature variations inside the spectrometer.
This model works in the eclipse period, when
the spacecraft experiences severe temperature variation
in the day -- night cycle, as well as in the day period.

Since our new method requires only spacecraft status data,
it is applicable to any spectrum recorded by the EIS.
It can be applied to both the short and the long wavelength bands,
as the instrumental effect produces the same amount of spectral drift
in the entire wavelength range of the EIS.
The correction method would be particularly useful for active region
and flare observations since no velocity reference is needed.
This procedure is provided as a part of the SSW.

\appendix   
\section{Spectral Curvatures of EIS}
Spectral curvatures on the detector are determined
by using long exposure data in quiet regions.
Correction of the curvatures is crucial for a precise
Doppler shift measurement with the EIS.
We selected two data sets recorded in a quiet region at disk center;
the first light of the EIS on 28 October 2006 and 
the velocity calibration data on 26 November 2009.

To cover the 1024-pixel height of the EIS detector,
two exposures in the top and bottom halves of the CCD
are required, since only a 512-pixel height
can be read out at one time.
Another exposure in the middle of detector
is needed to bridge them because
temperature variation of the spectrometer
causes instrumental spectral drift in the wavelength dimension
between the two exposures
(see main text for details).
The central position of the spectrum was determined by
fitting a Gaussian function at each slit height.
Fe~{\sc x} $\lambda$180.41~\AA, Fe~{\sc xii} $\lambda$195.12~\AA,
Fe~{\sc x} $\lambda$257.28~\AA, and S~{\sc x} $\lambda$264.23~\AA~
are selected as they provided good count statistics
and were not affected by blending lines.
The spectral drifts between the three exposures
were determined by comparing averaged spectral 
positions in overlapping regions.
The averaged velocities of common region were assumed to be
unchanged, although these spectra were recorded
at different times.
Spectra in the top and bottom halves were combined after
compensating the spectral drifts.

\begin{figure}
\centerline{\includegraphics[width=0.4\textwidth,clip=]{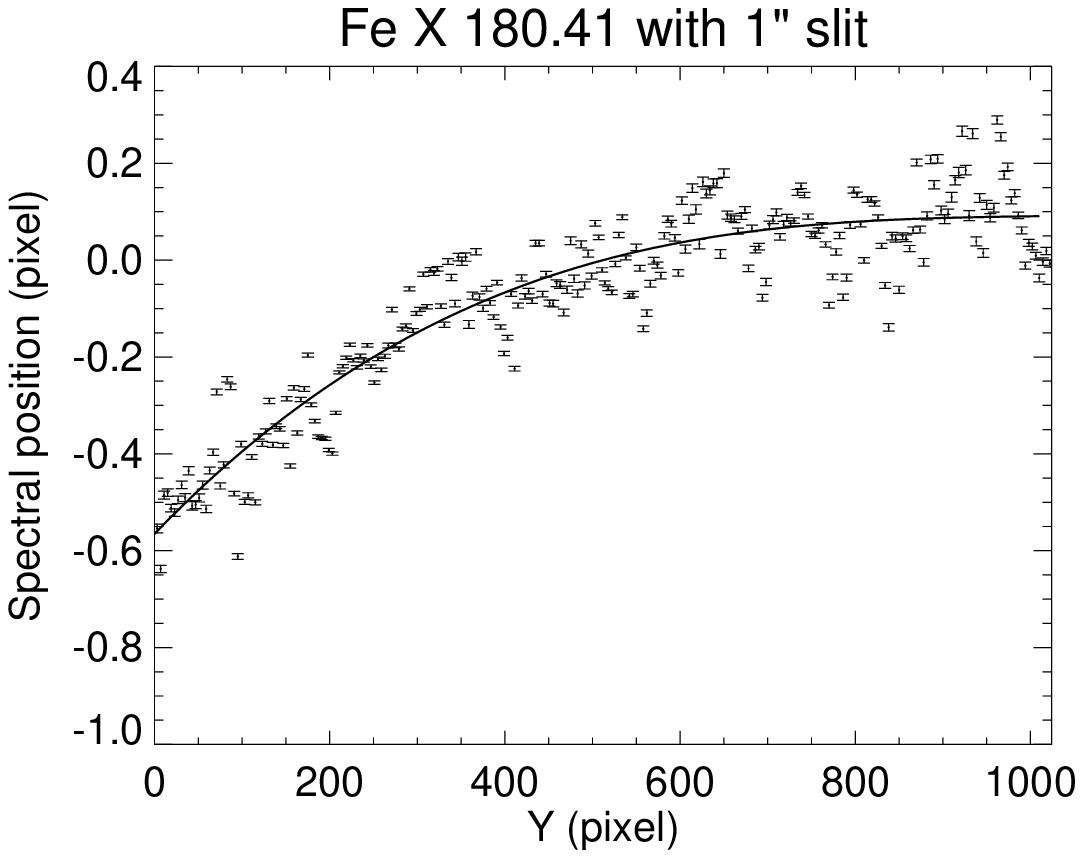}
\includegraphics[width=0.4\textwidth,clip=]{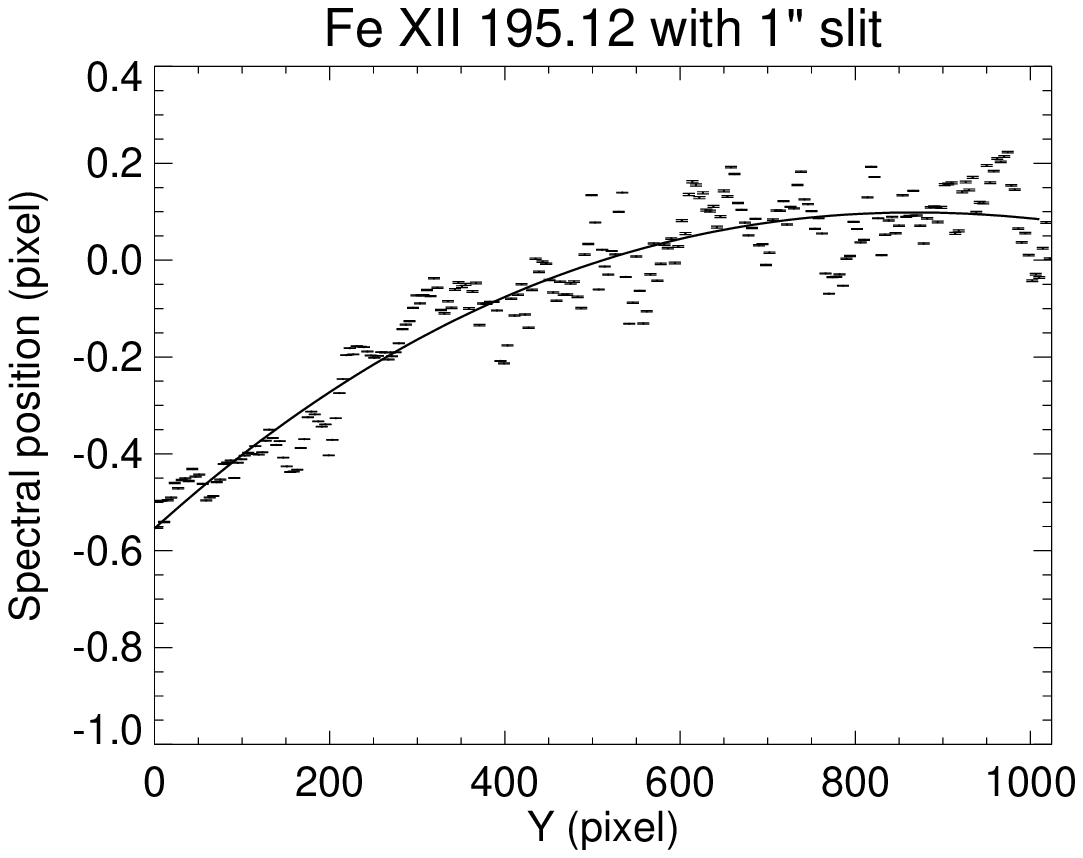}}
\caption{Spectrum center position determined from
Fe~{\sc x} $\lambda$180.41~\AA~ and
Fe~{\sc xii} $\lambda$195.12~\AA~ in the short wavelength band.
Positions are in pixel unit with respect to the detector center ($Y$=512).
Solid line presents the fitted third order polynomial function.}
\label{fig:slit:1s}
\end{figure}

Figure \ref{fig:slit:1s} presents spectral center
positions derived from the short wavelength band
data on 26 October 2006.
In order to smooth out small scale fluctuations,
they are fitted by a third order polynomial function.
The functions fitted to Fe~{\sc x} $\lambda$180.41~\AA~ and
Fe~{\sc xii} $\lambda$195.12~\AA~ are quite similar
and their difference is below the uncertainty of the
fitted functions.
It implies that the spectral curvatures are the same
within the short wavelength band.

\begin{figure}
\centerline{\includegraphics[width=0.4\textwidth,clip=]{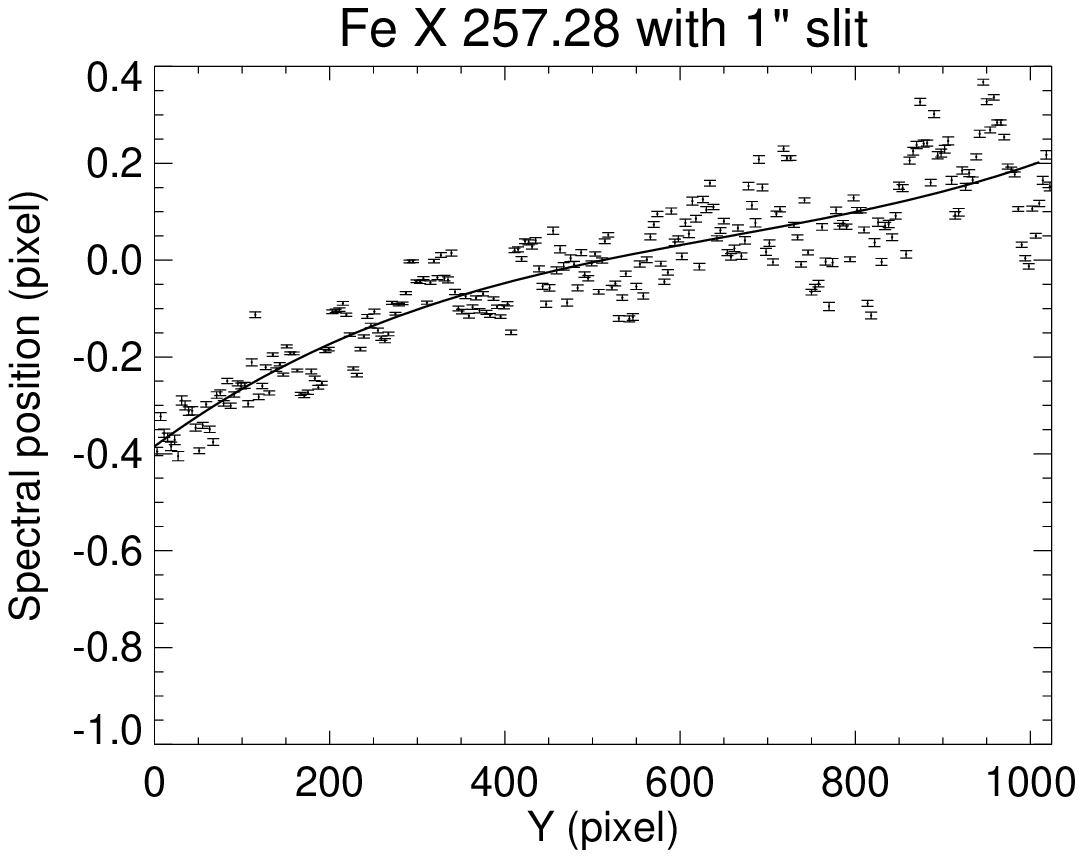}
\includegraphics[width=0.4\textwidth,clip=]{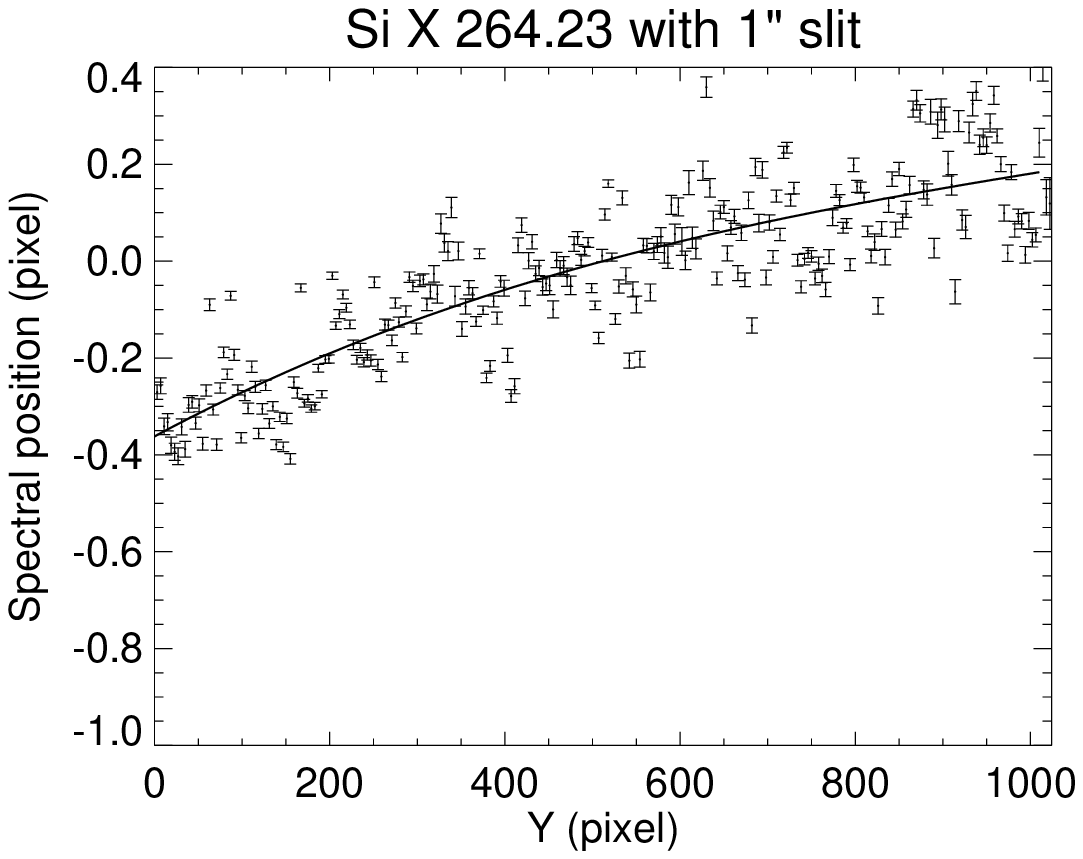}}
\caption{The same as Figure \ref{fig:slit:1s} but for
Fe~{\sc x} $\lambda$257.28~\AA~ and
S~{\sc x} $\lambda$264.23~\AA~ in the long wavelength band.}
\label{fig:slit:1l}
\end{figure}

Similarly, spectral curvatures in the long wavelength band
are determined from Fe~{\sc x} $\lambda$257.28~\AA~ and
S~{\sc x} $\lambda$264.23~\AA~ (Figure \ref{fig:slit:1l}).
Two spectral lines show very similar curvatures
though they are different from those of the short wavelength band.
To study the spectral curvatures in both wavelength bands
of EIS, at least one spectral line for each band must be analyzed.
In the following study,
Fe~{\sc xii} $\lambda$195.12~\AA~ in the short wavelength band and 
Fe~{\sc x} $\lambda$257.28~\AA~ in the long wavelength band are
analyzed since these lines had better count statistics than others.

\begin{figure}
\centerline{\includegraphics[width=0.8\textwidth]{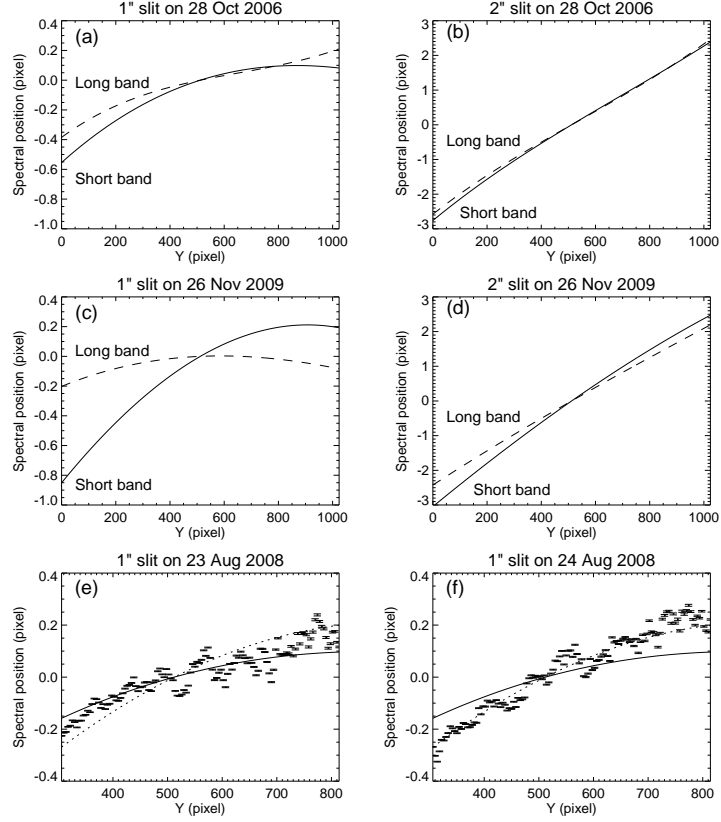}}
\caption{Variation of the EIS spectral curvature.
a: spectral curvatures with $1''$ width slit on 28 October 2006.
Solid line and dashed line respectively indicate the third order polynomial
functions fitted to Fe~{\sc xii} $\lambda$195.12~\AA~ in the short wavelength band
and Fe~{\sc x} $\lambda$ 257.28~\AA~ in the long wavelength band.
Positions are adjusted to zero at the detector center ($Y$=512).
b: the same as a but for $2''$ width slit.
c and d: the same format as a and b but for 26 November 2009
e: spectral positions of Fe~{\sc xii} $\lambda$195.12~\AA~ on 23 August 2008, before the
slit focus adjustment. Solid and dotted lines indicate
spectral curvatures on 28 October 2006 and on 26 November 2009,
respectively.
f: the same as e but for 24 August 2008, after the slit focus
adjustment.
}
\label{fig:slit:plot}
\end{figure}

Figures \ref{fig:slit:plot}a and \ref{fig:slit:plot}b present functions fitted
to spectra on 28 October 2006.
The long and the short wavelength bands show different
functions as mentioned above.
The $2''$ width slit has larger tilt with respect to the detector than
the $1''$ width slit and causes a large linear slope of the spectrum.
Spectral curvatures on 26 November 2009 show quite different
characteristics (Figures \ref{fig:slit:plot}c and \ref{fig:slit:plot}d).
Compared to 28 October 2006 in Figures \ref{fig:slit:plot}a and \ref{fig:slit:plot}b,
the tilt of both slits increased in the short wavelength band,
while they decreased in the long wavelength band.

A possible cause of the spectral curvature change is
movement of the slit and the grating during
the period.
Figures \ref{fig:slit:plot}e and \ref{fig:slit:plot}f present offlimb spectra
before and after the slit focus adjustment on 24 August 2008.
Although it covers only $512''$ height along the slit,
the effect of the the slit focus adjustment can be assessed by comparing them.
In Figure \ref{fig:slit:plot}e before the slit adjustment,
spectral positions fall on the solid line of 28 October 2006.
After the slit adjustment (in Figure \ref{fig:slit:plot}f),
it fits well on the dashed line of 26 November 2009.
The results suggest that the slit focus movement
on 24 August 2008 is the primary cause of spectral
curvature change. 
Although the grating was moved on 21 October 2008,
it seems only to have a minor impact on spectral curvatures.
No other movement of the optical element of the EIS has been done
by the time this article was written.

\begin{table}
\caption{Fitted parameters for spectral curvatures}
\label{table:slit}
\begin{tabular}{ccccccc} 
\hline                   
 Date & Slit & Band & $c_0$ & $c_1$ & $c_2$ & $c_3$ \\
\hline
28 Oct  2006 & 1'' & short &
$-0.55$ & $1.63\times 10^{-3}$ &$-1.17\times 10^{-6}$ &$1.73\times 10^{-10}$\\
& & long &
$-0.38$ & $1.33\times 10^{-3}$ &$-1.54\times 10^{-6}$ &$7.87\times 10^{-10}$\\
& 2'' & short &
$-2.76$ & $6.44\times 10^{-3}$ &$-2.71\times 10^{-6}$ &$1.30\times 10^{-9}$\\
& & long &
$-2.58$ & $6.15\times 10^{-3}$ &$-3.15\times 10^{-6}$ &$1.92\times 10^{-9}$\\
26 Nov 2009 & 1'' & short &
$-0.84$ & $2.19\times 10^{-3}$ &$-9.76\times 10^{-7}$ &$-1.75\times 10^{-10}$\\
& & long &
$-0.18$ & $7.39\times 10^{-4}$ &$-9.00\times 10^{-7}$ &$2.94\times 10^{-10}$\\
& 2'' & short &
$-3.04$ & $6.40\times 10^{-3}$ &$-8.01\times 10^{-7}$ &$-1.88\times 10^{-10}$\\
& & long &
$-2.42$ & $5.08\times 10^{-3}$ &$-7.94\times 10^{-7}$ &$2.32\times 10^{-10}$\\
\hline
\end{tabular}
\end{table}
Parameters for the fitted functions are summarized in Table
\ref{table:slit}.
The spectral shift $s$ at given slit height is expressed as
\begin{equation}
s = c_0 + c_1 y + c_2 y^2 + c_3 y^3
\end{equation}
where $y$ is a position in CCD pixel coordinates. 
Before the slit adjustment on 24 August 2008,
spectral curvatures can be corrected by using
the parameters fitted to spectra on 28 October 2006.
After the slit adjustment, the spectral curvatures can be
compensated for by using another set of parameters
fitted to spectra on 26 November 2009.

\section{Spatial Offset of the Spectrum}
Because of the tilt of the grating with respect to the detector,
spectra recorded by EIS have a wavelength dependent offset
in the north -- south direction \cite{young2009}.
To compare multiple emissions from the same feature of the Sun,
it is essential to correct the spatial offset along the slit.
The Mercury transit observed by the EIS on 8 November 2006
provided a unique opportunity to determine
the spatial offset.
Since the solar emission is obscured by Mercury,
a shadow must be seen at identical locations in all wavelengths.
By using the EIS slot observing mode, two dimensional images
in strong emission lines were recorded.
The spatial offset is determined
by measuring the positions of the Mercury shadow in selected emission lines.

\begin{figure}
\centerline{\includegraphics[width=0.5\textwidth,clip=]{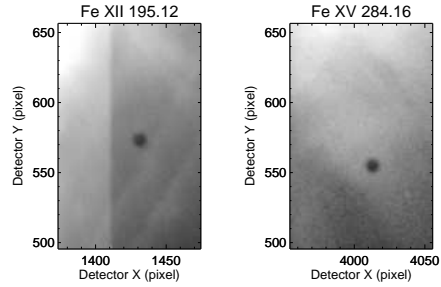}}
\caption{Left: Mercury transit seen in Fe~{\sc xii}~$\lambda$195.12~\AA~ emission, displayed in log scale. Axis indicate dimensions in CCD pixel.
Right: Emission in Fe~{\sc xv} $\lambda$284.16~\AA.}
\label{fig:offset:slot}
\end{figure}

Figure \ref{fig:offset:slot} shows images during the Mercury transit
in Fe~{\sc xii} $\lambda$195.12~\AA~ and Fe~{\sc xv} $\lambda$284.16~\AA,
which were obtained at 18:52 UTC by the $266''$-wide slot.
Mercury produced a round shadow against
a bright emission from the solar corona.
The apparent diameter of Mercury was about 10 pixel on the
detector, where the plate scale is $1''$ pixel$^{-1}$.
Other images were obtained with the $40''$-wide slot later at 21:30 UTC.
Since two dimensional images obtained with the EIS slots are overlapping
with nearby emission lines,
prominent emission lines are selected for this study, namely;
Fe~{\sc xi} $\lambda$181.23~\AA,
Fe~{\sc xii} $\lambda$193.51~\AA,
Fe~{\sc xii} $\lambda$195.12~\AA,
Fe~{\sc xiii} $\lambda$202.04~\AA,
He~{\sc ii} $\lambda$256.32~\AA,
Fe~{\sc xiv} $\lambda$264.79~\AA,
Fe~{\sc xiv} $\lambda$274.20~\AA, and
Fe~{\sc xv} $\lambda$284.16~\AA.
The $40''$ slot observations allowed measurements in more lines
than did the $266''$ slot
as the overlap of emission lines is limited.
The position of Mercury at each emission line is
determined by measuring the centroid of the shadow
profile in the north -- south direction.
The estimated accuracy of the shadow position measurement is 0.3 pixel.

\begin{figure}
\centerline{\includegraphics[width=0.8\textwidth,clip=]{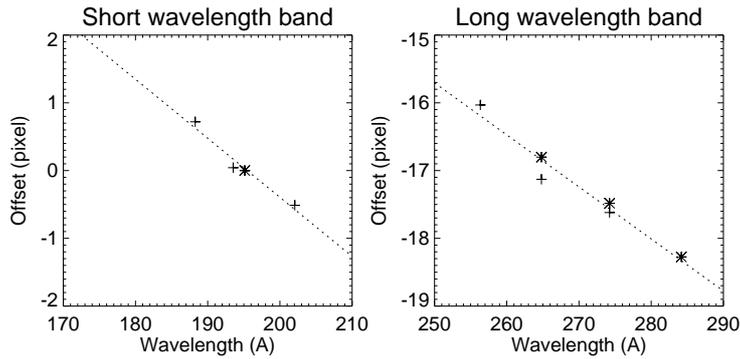}}
\caption{Left: Spatial offsets measured in the short wavelength
band from the Mercury transit on 8 November 2006.
Plus and asterisk signs indicate offsets deduced from
$40''$-wide slot and $266''$-wide slot images of the EIS.
The offset at Fe~{\sc xii} $\lambda$195.12~\AA~ is adjusted to zero.
Right: Spatial offsets in the long wavelength band with
respect to Fe~{\sc xii} $\lambda$195.12~\AA.}
\label{fig:offset:plot}
\end{figure}

Figure \ref{fig:offset:plot}
presents spatial offsets measured at different wavelength.
Measurements from the $40''$ and $266''$ wide slots
give a consistent spatial offset.
A linear function fitted to the short wavelength band is
expressed as
\begin{equation}
y = -8.72 \times 10^{-2} \lambda + 17.01
\label{eq:offset:sw}
\end{equation}
where $\lambda$ is the wavelength in \AA.
\inlinecite{young2009}
determined a gradient of $-7.92 \times 10^{-2}$
in the short wavelength band.
Their value is consistent with our result since
the discrepancy within the commonly used range of
180~\AA~ to 203~\AA~ is only 0.18 pixel,
which is close to the uncertainty of the offset
measurement.
The spatial offset for the long wavelength band is
determined in the same manner.
\begin{equation}
y = -7.68 \times 10^{-2} \lambda + 3.49
\label{eq:offset:lw}
\end{equation}
The spatial offset in both wavelength bands of
the EIS can be compensated for by using
the plate scale of $1.002\pm0.016$ arcsec pixel$^{-1}$
\cite{hara2008b} and equations (\ref{eq:offset:sw}) and (\ref{eq:offset:lw}).
In addition, \inlinecite{hara2008b} also determined
the normal scan step size of $0.961\pm0.004$ arcsec,
which can be used to deduce the width of a raster scan
in east -- west direction.

%
\begin{acks}

{\it Hinode} is a Japanese mission developed and launched by ISAS/JAXA, with NAOJ as domestic partner and NASA and STFC (UK) as international partners.
It is operated by these agencies in co-operation with ESA and NSC (Norway). 
Authors are grateful to the ROOT development team for
providing a useful software package. We would like to thank
Dr. J. M. Borrero for fruitful discussion on developing artificial neural network.

\end{acks}
\bibliographystyle{spr-mp-sola}
\bibliography{reference.bib}

\end{article}
\end{document}